\documentclass[preprint,showpacs,preprintnumbers,amsmath,amssymb]{revtex4}

\usepackage[dvips]{graphicx}
\usepackage{subfigure}

\begin{document}

\title{Effects of homogenous broadening on the Rabi splitting in micropillar
cavities with strong light-matter interaction}

\author{L. M. Le\'on Hilario}

\affiliation{Facultad de Ciencias, Universidad Nacional de Ingenier\'{\i}a, Per\'{u}}

\author{A. A. Aligia}

\affiliation{Centro At\'{o}mico Bariloche and Instituto Balseiro, Comisi\'{o}n Nacional
de Energ\'{\i}a At\'{o}mica, 8400 Bariloche, Argentina}

\pacs{78.67.Hc, 78.55.-m, 71.36.+c, 42.50.Pq}

\begin{abstract}
We solve the low-energy part of the spectrum of a model that describes a cavity mode
strongly coupled to an exciton, and both modes coupled to continua of bosonic excitations
which give rise to homogeneous broadenig. The spectral density of the cavity modes in 
the low-energy manifold agrees with measured photoluminiscense spectra. We suggest fitting 
these spectra with a sum of two asymmetric Lorentzians.
\end{abstract}

\maketitle

\section{Introduction and motivation}

In the last years there has been great interest in the field of cavity
quantum electrodynamics. In particular, systems with strong coupling between
single quantum dots (QDs) and high quality microcavities have been studied
for different reasons, such us to gain insight into different quantum optics
effects \cite{rei,yos,press,reit,dal,optic,mich,ima}, like quantum
decoherence, entanglement and possible applications in quantum information
processing \cite{rei,yos,press,reit}. For example, some of these systems were
proposed as a single-photon source \cite{press,mich} for realization of all
optical quantum computing \cite{ima}. The SC regime takes place when the
coupling between a single quantum emitter and cavity mode is stronger
compared to their decay rates. In this case, the emitter and cavity
coherently exchange energy back and forth leading to Rabi oscillations. The
SC between single (In,Ga)As QD and micropillar cavity modes \cite{rei}, has
become apparent in photoluminiscence data which displayed anti-crossings
between the QD exciton and cavity-mode dispersion relations \cite%
{rei,yos,reit}. While usually temperature was used to tune the energy of the
excitonic transition to that of the cavity mode, it was shown recently that
the magnetic field can also be used as a tuning parameter\cite{reit}.

Some of the experimental works \cite{rei,press,reit} analyze their data in
terms of a 2x2 matrix that mixes the exciton and the cavity mode. To
introduce life time effects, complex energies are used to represent the
energies of the uncoupled system and therefore, the matrix is non Hermitian.
Similar expressions for the Rabi splitting were obtained using a master
equation within a phenomenological framework \cite{car,andre}, but to our
knowledge, a microscopic description of the system which includes finite
life times of the exciton and the cavity mode is still lacking,

In this paper, we extend the Hamiltonian which describe the coupling of the
exciton and cavity modes \cite{andre} to include the broadening of both
excitations, due to a mixture with a continuum of bosonic excitations. The
problem can be solved rigorously for weak excitation. We compare our results
for the intensity of photoluminiscence with recent experiments\cite%
{rei,reit}.


\section{Model}

$_{{}}$\label{Model} 
The core of the model contains the cavity photon mode, the excitonic degrees
of freedom represented by a spin $1/2$, and the coupling between them \cite%
{andre}. We include the coupling of the cavity mode with a continuum of
radiative modes which gives rise to the broadening of the cavity mode (the
most important one) \cite{leon1,leon2,bruc}. We also couple the excitonic
mode with a continuum of bosonic excitations, leading to a broadening of the
excitonic energy. The Hamiltonian is 
\begin{eqnarray}
H &=&E_{x}S_{z}+E_{c}a^{\dagger }a+(VS^{-}a^{\dagger }+\text{H.c.}%
)+\sum_{r}\epsilon _{r}a_{r}^{\dagger }a_{r}  
+\sum_{r}(V_{r}a_{r}^{\dagger }a+\text{H.c.})+\sum_{\nu }\epsilon _{\nu
}b_{\nu }^{\dagger }b_{\nu }  \nonumber \\
&&+\sum_{\nu }(V_{\nu }b_{\nu }^{\dagger }S^{-}+\text{H.c.}).  \label{mod}
\end{eqnarray}%
where $a^{\dagger }$ is the creation operator of the cavity mode, $%
S_{z},S^{+},S^{-}$ are spin operators for the two level system of the QD
with ground $(|\downarrow \rangle )$ and excited $(|\uparrow \rangle )$
state which represent zero and one exciton respectively, $a_{r}^{\dagger }$
creates the radiative mode $r$ which couples to the cavity mode and
similarly $b_{\nu }^{\dagger }$ creates a bosonic excitation $\nu $ coupled
to the exciton. For simplicity, the subscripts indicating the polarization
of the modes are dropped.

The first three terms of the Hamiltonian describe the strong coupling
between the cavity mode and the exciton \cite{andre}. The fourth and fifth
terms describe a continuum of radiative modes and its coupling to the cavity
mode. The following two terms have a similar effect for the exciton mode.

The model is similar to the one previously used by us to describe Raman
experiments in microcavities with quantum wells inside them \cite%
{leon1,leon2,bruc}. The main difference is that in the previous case, the
problem has a two dimensional translational symmetry leading to delocalized
excitons, and for each wave vector the probability of occupation of the
excitonic state is very small. This allows to treat the excitons as bosonic
excitations with a high degree of accuracy \cite{leon1}. This is not possible
in the present case and the model becomes highly non trivial. In spite of
this, some exact results can be derived, using the fact that the total
number of excitations

\begin{equation}
N_{e}=S_{z}+1/2+a^{\dagger }a+\sum_{r}a_{r}^{\dagger }a_{r}+\sum_{\nu
}b_{\nu }^{\dagger }b_{\nu }  \label{ne}
\end{equation}
is conserved. For example, clearly the ground state is the only state with $%
N_{e}=0$.

In the following,we denote by $|n,S_{z}\rangle $ the states of the system
with $n$ cavity photons, exciton state $S_{z}$ and no bosons described by $%
a_{r}$ or $b_{\nu }$. If one of the latter is occupied, we denote the
corresponding state by $|n,S_{z};r\rangle $ or $|n,S_{z};\nu \rangle $. The
Hilbert subspace with $N_{e}=1$ can be treated analytically. This subspace
consist of a state formed by zero cavity photon and one exciton, $%
|0,\uparrow \rangle $, one cavity mode (and no excitons) $|1,\downarrow
\rangle $, and two continua of bosonic excitations $|0,\downarrow ;r\rangle $%
, $|0,\downarrow ;\nu \rangle $. The problem in this subspace takes a form
similar to that of an impurity interacting with a continuum or the resonant
level model.

The photoluminscence measurements in micropillar cavities \cite{rei,yos,reit}
suggest that the only subspace relevant for the experiments is that with $%
N_{e}=1$. To see this, let us neglect for the moment the modes $r$ and $\nu $
responsible for the broadening of the cavity mode and the exciton. The
resulting model which contains the three first terms of Eq. (\ref{mod}) can
be solved exactly \cite{andre}. For each subspace of definite $N_{e}$, the
model consists of two branches which have an anticrossing as a function of
the detuning $\Delta =E_{x}-E_{c}$. The value of the Rabi splitting is $2V%
\sqrt{N_{e}}$. Experimentally only one anticrossing is reported. This
indicates that the probability of exciting two modes is low and that the
theoretical results for $N_{e}=1$ are enough to describe the experiments.

\section{Green's functions}

Motivated from the above discussion, we assume that the observed
photoluminiscence is proportional to the density of cavity photons in the
subspace with $N_{e}=1$. In other words, the observed intensity is
proportional to the density

\begin{equation}
\rho _{11}(\omega )=-\frac{1}{\pi }\mathrm{Im}G_{11}(\omega ),  \label{rho}
\end{equation}%
where $G_{11}(\omega )$ is a many-body Green's function obtained from the
diagonal matrix element of the state $|1,\downarrow \rangle $ (for brevity
we denote $|1\rangle \equiv |1,\downarrow \rangle, |0\rangle \equiv
|0,\uparrow \rangle, |r\rangle \equiv |0,\downarrow;r \rangle,|\nu\rangle
\equiv |0,\downarrow;\nu \rangle$) of the operator given by the inverse of
the Hamiltonian in the subspace $N_{e}=1:$

\begin{equation}
G_{11}(\omega )=\langle 1|G(\omega )|1\rangle ;\text{ }G(\omega )=(\omega
+i0^{+}-H)^{-1}.  \label{g11}
\end{equation}%
The matrix elements of the operator $G(\omega )$ can be evaluated from the
equation $G(\omega -H)=I$, where $I$ is the identity matrix. Taking the
element $\left\lbrace 11 \right\rbrace $ of this equation one obtains 
\begin{equation}
G_{11}(\omega )\left( \omega -E_{c}\right) -VG_{10}(\omega
)-\sum_{r}V_{r}G_{1r}(\omega )=1.  \label{ecg1}
\end{equation}%
Proceeding in a similar way for matrix elements $\left\lbrace 10 \right\rbrace , \left\lbrace 1r \right\rbrace$, 
and the indices of
the new Green%
\'{}%
s functions that appear, the system of equations for $G_{ij}$ can be closed.
In particular the relations 
\begin{eqnarray}
G_{10}(\omega ) &=&\frac{V_{r}G_{11}}{(\omega -E_{x}-S_{x})}  \nonumber \\
G_{1r}(\omega ) &=&\frac{V_{r}}{(\omega -\epsilon _{r})}G_{11},  \label{ecg2}
\end{eqnarray}%
Replaced in Eq. (\ref{ecg1}) gives a closed expression for the Green%
\'{}%
s function $G_{11}$: 
\begin{equation}  
G_{11}(\omega )=\frac{1}{\omega -E_{c}-S_{c}-\frac{V^{2}}{\omega -E_{x}-S_{x}%
}},  \label{green}
\end{equation}%
where 
\begin{eqnarray}
S_{c}(\omega ) &=&\sum_{r}\frac{V_{r}^{2}}{\omega +i0^{+}-\epsilon _{r}}\ , 
\nonumber \\
S_{x}(\omega ) &=&\sum_{\nu }\frac{V_{\nu }^{2}}{\omega +i0^{+}-\epsilon
_{\nu }}.  \label{sums}
\end{eqnarray}%
In this paper, for simplicity we assume (as usual in related problems)
constant $V_{r}$, $V_{\nu }$ and densities of the modes $r$ and $\nu $, near
the point of zero detuning $\Delta =E_{x}-E_{c}$. Then, except for some real
shifts that can be incorporated in $E_{x}$ and $E_{c}$ \cite{leon1}, the
above sums reduce to imaginary constants that we take as parameters 
\begin{equation}
S_{c}(\omega )=-i\delta _{c}\ ,\qquad S_{x}(\omega )=-i\delta _{x}\ .
\label{delta}
\end{equation}

Note that interchanging the subscripts $x$ and $c$, Eqs. (\ref{green}) and (
\ref{delta}) give the density of the excitonic excitation $|0,\uparrow
\rangle $.


\section{Analysis and discussion}

\label{Analisys and discussion} 

\subsection{The density as sum of two asymmetric Lorentzians}

With a straightforward manipulation, Eq. (\ref{green}) can be written a sum
of two fractions, with denominators linear in $\omega $. Replacing in Eq. (%
\ref{rho}), the density  $\rho _{11}$ can be separated as 
\begin{eqnarray}
\rho _{11}(\omega ) &=&\rho _{11}^{1}(\omega )+\rho _{11}^{2}(\omega ), 
\nonumber \\
\rho _{11}^{1}(\omega ) &=&\frac{1}{\pi }\frac{\delta _{1}a_{1}+d(\omega
-\omega _{1})}{(\omega -\omega _{1})^{2}+\delta _{1}^{2}},  \nonumber \\
\rho _{11}^{2}(\omega ) &=&\frac{1}{\pi }\frac{\delta _{2}a_{2}-d(\omega
-\omega _{2})}{(\omega -\omega _{2})^{2}+\delta _{2}^{2}},  \label{dens}
\end{eqnarray}%
where 
\begin{eqnarray}
a_{1(2)} &=&\frac{1}{2}\pm \frac{x\Delta -y\delta }{4(x^{2}+y^{2})} 
\nonumber \\
\omega _{1(2)} &=&c\pm x  \nonumber \\
\delta _{2(1)} &=&c^{\prime }\pm y  \nonumber \\
d &=&\frac{y\Delta -x\delta }{4(x^{2}+y^{2})}  \label{dens2}
\end{eqnarray}%
with 
\begin{eqnarray}
x &=&\frac{1}{2}\sqrt{A+\frac{1}{2}\sqrt{B}}  \nonumber \\
y &=&\frac{\Delta \delta }{4x}  \nonumber \\
c &=&\frac{E_{c}+E_{x}}{2}  \nonumber \\
c^{\prime } &=&\frac{\delta _{c}+\delta _{x}}{2}  \label{xy}
\end{eqnarray}%
and 
\begin{eqnarray}
A &=&\frac{\Delta ^{2}-\delta ^{2}+4V^{2}}{2}  \nonumber \\
B &=&(\Delta ^{2}+\delta ^{2})^{2}+8V^{2}(\Delta ^{2}-\delta ^{2})+16V^{4} 
\nonumber \\
\delta  &=&\delta _{c}-\delta _{x}.  \label{ab}
\end{eqnarray}

The result can be interpreted as the sum of two asymmetric Lorentzians, with
opposite asymmetries controlled by $d$. Neglecting the effect of $d$, the
position, amplitude and width of the Lorentzians is given by $\omega _{i}$, $%
a_{i}$ and $\delta _{i}$ respectively with $\omega _{1}<\omega _{2}$ except
for detuning $\Delta =E_{x}-E_{c}=0$ if $2V\leq \delta $. The results for $%
\omega _{i}$, and $\delta _{i}$ agree with the real and imaginary parts of
the complex roots of a $2\times 2$ matrix, given by Eq. (1) of Press et al%
 \cite{press}. The above results provide a microscopic justification for this
expression.

It is easy to see from Eqs. (\ref{dens})-(\ref{ab}) that in the limit $%
\delta _{c}\longrightarrow \delta _{x}$, $d\longrightarrow 0$, and therefore
the density is give by the sum of two Lorentzians separated $2x$ with $x=%
\sqrt{((E_{c}+E_{x})/2)^{2}+V^{2}}$ and with widths $\delta _{c}$ and $%
\delta _{x}$

\begin{figure}[tbp]
\includegraphics[width=10.5cm]{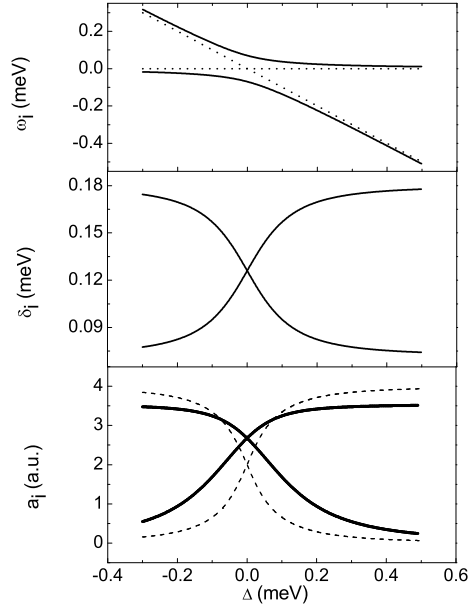}
\caption{Peak energies, linewidths and intensity in function of detuning for
the parameters of Ref. \onlinecite{rei}: $\delta_c=0.09$ meV, $\delta_x=0.036$ meV, $V=0.075$ meV.  
The dashed line correspond to the maxima of $\rho_{11}(\omega)$.}
\label{lor}
\end{figure}

In Fig. \ref{lor}, we show the resulting parameters of the two peaks for the
experiment of Ref. \onlinecite{rei}. There is a qualitative agreement with Fig. 4
of Ref. \onlinecite{rei} 

\begin{figure}[tbp]
\includegraphics[width=8.5cm]{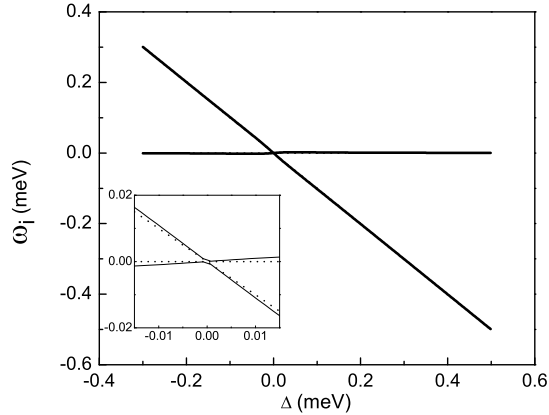}
\caption{Peak energies for the case of weak coupling. Parameters are 
$\delta_c=0.09$ meV, $\delta_x=0.036$ meV, $V=0.022$ meV.}
\label{wea}
\end{figure}

In Fig. \ref{wea}, we show the position of the peaks  $\omega _{i}$ for a case in
which $V$ was reduced to $V=0.022$ meV, so that there is no splitting of the
energies for zero detuning.

\subsection{Position of the maxima}

When $V\gg \delta _{c},\delta _{x}$, the spectral density $\rho _{11}(\omega
)$ shows two maxima for all values of the detuning $\Delta $. The position
of the maxima are given by some real roots of a polynomial of fifth degree
obtained from the condition $\partial \rho _{11}(\omega )/\partial \omega =0$%
. As $V$ decreases, the two maxima merge into one for zero detuning, at a
critical value $V_{c}$.

\begin{figure}[tbp]
\includegraphics[width=8.5cm]{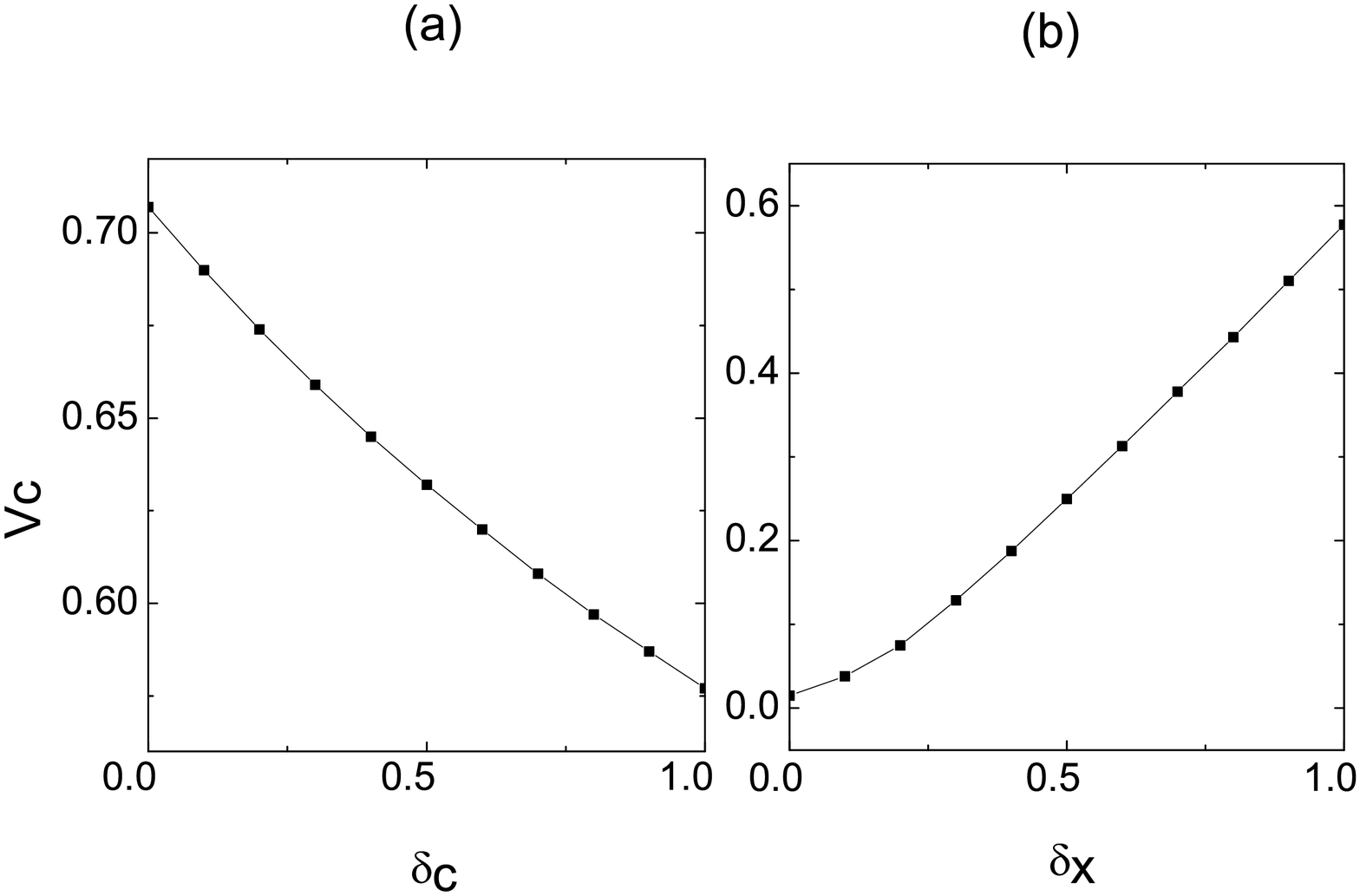}
\caption{ Critical coupling as a function of (a) linewidth of the cavity mode 
$\protect\delta _{c}$, and (b) linewidth of exciton mode $\protect\delta _{x}$%
}
\label{vc}
\end{figure}

In Fig. \ref{vc} we represent $V_{c}$ as a function of one of the
broadenings $\delta _{c}$ or $\delta _{x}$, keeping constant the other one.
We can that as $\delta _{x}$ increases, $V_{c}$ also increases. However, the
trend is the opposite as $\delta _{c}$ increases. 

\begin{figure}[tbp]
\includegraphics[width=8.5cm]{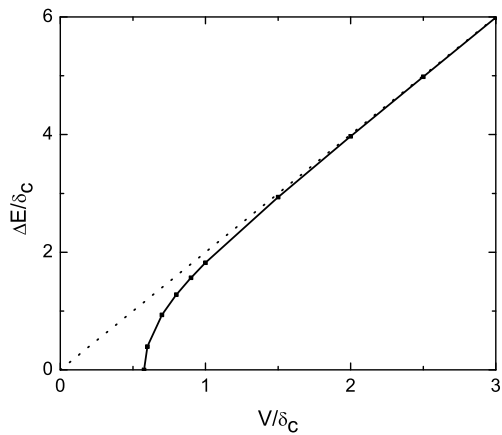}
\caption{Separation between the maxima of $\rho_{11}(\omega)$
as a function of coupling contant $V$ for 
$\protect\delta _{x}=\protect\delta _{c}$ and $E_{x}=E_{c}$}
\label{de1}
\end{figure}

In Fig. \ref{de1} we plot the difference between the two maxima $\ \Delta E$
as function of $V$ for the case $\delta _{x}=\delta _{c}$. The
experimentally measured Rabi splitting might be identified with $\ \Delta E$%
, but another possibility is to relate the Rabi splitting with $\omega
_{2}-\omega _{1}=2x$ [see Eqs. (\ref{dens})-(\ref{ab})]. We believe that the
latter approach is more physical if the experimental line can be fit by Eq. (%
\ref{dens}).

We can see from the figure that $\Delta E$ has a behavior of a square root
as a function of $\Delta V=V-V_{c}$ for small $\Delta V$, while for large $V$%
, $\Delta E$ is proportional to $2V$, as expected. 


\subsection{Spectral density for different detunings}

\begin{figure}[tbp]
\includegraphics[width=9.0cm]{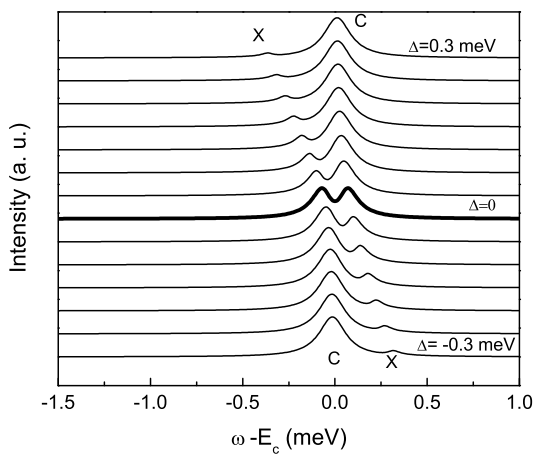}
\caption{Photoluminescense spectra for several values of detunning 
$\Delta =E_{x}-E_{c}$. The parameter used are
the same as in Fig. \ref{lor} \cite{rei}.}
\label{inte}
\end{figure}

The spectral density (assumed proportional to the photoluminiscence
intensity) for different detuning $\Delta =E_{x}-E_{c}$, is presented in
Fig. \ref{inte} for \ parameters that correspond to \ the particular
experimental work of Reithmaier \textit{et al. }\cite{rei} in which the
detuning was controlled by the tempertaure The anticrossing is clearly
visible and the variation of the intensity profile is similar to the
observed one. For negative detuning, the peak at lowest energy is more
excitonic like and therefore has lower intensity than the other one, which
has a greater admixture with the cavity mode. The situation is reversed for
positive detuning.

\begin{figure}[tbp]
\includegraphics[width=9.5cm]{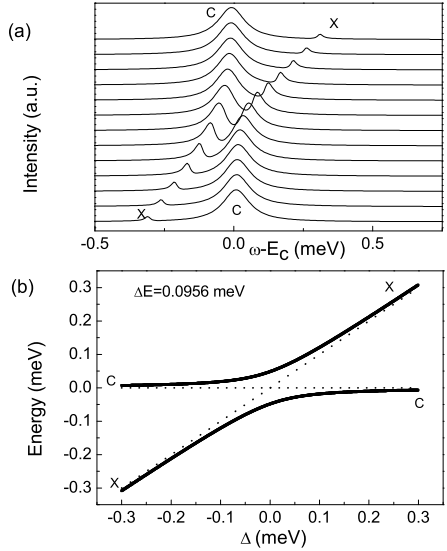}
\caption{(a) Photoluminescense spectra for several values of detuning,
(b) Energy dispersion of the two emission modes.
the uncoupled modes is indicated by dot lines.
Parameters are $\delta_c=0.06$ meV, $\delta_x=0.01$ meV, $V=0.046$ meV}
\label{intep}
\end{figure}

In Ref. \onlinecite{reit}, the detuning was controlled by the application of a
magnetic field. In Fig. \ref{intep} we show the corresponding theoretical
curve, with the same qualitative trends as before. The coupling $V=0.046$
meV was adjusted in such a way that the difference between the maxima in $%
\rho _{11}(\omega )$ correspond to the reported Rabi splitting of $0.0956$ meV
and half width at half maximum of the cavity and exciton modes $\delta
_{c}=0.06$ meV and $\delta _{x}=0.01$ meV respectively (corresponding to
half of the reported full width at half maximum). 

\section{Summary}

\label{Conclusions} 

We have studied a microscopic model that couples a cavity mode with an
exciton, and includes coupling to two continua of bosonic excitations, which
give rise to a homogeneous broadening of both modes. Although the model is
not exactly solvable, we treat exactly the low-energy spectrum and provide
expressions for the low-energy part of the spectral density of the cavity
mode and the exciton. The former agrees with measured photoluminiscence
spectra for several detunings.

Our approach provides a microscopic justification for simple
phenomenological expressions for the position and widths of the two mixed
modes, between the cavity mode and the exciton, when both modes have a
homogeneous broadening.

\section*{Acknowledgments}

We thank CONICET from Argentina for financial support. This work was
partially supported by PIP 11220080101821 of CONICET, and PICT 2006/483 and
PICT R1776 of the ANPCyT.

\end{document}